\documentstyle[aps,array]{revtex}

\newcommand{\be}[1]{\begin{equation} \label{(#1)}}
\newcommand{\ee}{\end{equation}}

\begin{document}

\author{Thomas W. Kephart and Heinrich P\"{a}s}
\title{Three Family ${\cal N}=1$ $SUSY$ Models from $Z_{n}$ Orbifolded
$AdS/CFT$}
\date{13 September 2001}
\address{Department of Physics and Astronomy,\\
Vanderbilt University, Nashville, TN 37325.}
\maketitle

\bigskip

\begin{abstract}

We present an analysis of compactifications of the type $IIB$
superstring on
$AdS_{5}\times S^{5}/\Gamma $, where $\Gamma $ is an abelian cyclic
group.
Every $\Gamma =Z_{n}$ of order $n\leq 12$ is considered. This results in
60
chiral models, and a systematic analysis with $n<8$
yields
four containing the minimal $SUSY$ standard model with three families.
One
of these models extends to an infinite sequence of three-family $MSSMs$.
We also give a lower bound on the number of chiral models for all values
of $n$.

\end{abstract}


\bigskip

\preprint{%
\vbox{\hbox {June 2000} \hbox{IFP-784-UNC}
\hbox{VAND-TH-00-2} \hbox{hep-th/0109111}}}

\newpage


Orbifolded $AdS_{5}\times S^{5}$ is fertile ground for building models
which
can potentially test string theory. When one
bases the models on the conformal field theory gotten from
the large $N$ expansion of the $AdS/CFT$ correspondence \cite{Maldacena:1998re},
stringy effects
can
show up at the scale of a few $TeV$. The first three-family model of
this
type had ${\cal N}=1$ $SUSY$ and was based on a $Z_{3}$ orbifold \cite{Kachru:1998ys}. However, since
then
the most studied examples have been models without supersymmetry based
on
both abelian \cite{Frampton:1999wz}, \cite{Frampton:1999nb}, \cite{Frampton:1999ti} and non-abelian \cite{Frampton:2000zy}, \cite{Frampton:2000mq} orbifolds of $AdS_{5}\times S^{5}$. Here we
return to $Z_{n}$ orbifolds with supersymmetry, and systematically study
those cases with chiral matter ($i. e.$, those with an imbalance between chiral supermultiplets and
anti-chiral
supermultiplets). We classify all cases up
to $n\leq 12$, and show that several of these contain the minimal
supersymmetric standard model ($MSSM$) with three families.

\bigskip

Let us summarize how these models are constructed (for a review see \cite{Frampton:2000mq}). First we choose a discrete
group $\Gamma $ with which to form the orbifold $AdS_{5}\times
S^{5}/\Gamma $%
. The replacement of $S^{5}$ by $S^{5}/\Gamma $ reduces the
supersymmetry to
${\cal N}=$ 0, 1 or 2 from the initial ${\cal N}=4$, depending on how $%
\Gamma $ is embedded in the $SO(6) \sim SU(4)$ isometry of
$S^{5}$.
The case of interest here is ${\cal N}=1$ $SUSY$ where $\Gamma $
completely
embeds in an $SU(3)$ subgroup of the $SU(4)$ isometry. $I.e$., we embed
rep($%
\Gamma )\rightarrow {\bf 4}$ of $SU(4)$ as ${\bf 4}=({\bf 1},{\bf r})$
where
${\bf 1}\ $is the trivial irrep of $\Gamma $ and ${\bf r}$ is a
nontrivial
(but possibly reducible) three dimensional representation of $\Gamma .$
The
chiral supermultiples generated by this embedding are given by
\be{}
\sum_{i}{\bf 4}\otimes R_{i}
\ee
where the set \{$R_{i}\}$ runs over all the irreps of $\Gamma .$ For our
choice, $\Gamma =Z_{n},$ the irreps are all one dimensional
and as a consequence of the choice of $N$ in the $1/N$ expansion,
the gauge group \cite{Lawrence:1998ja} is $SU^{n}(N)$. Chiral models require the {\bf 4} to be
complex (${\bf 4}\neq {\bf 4}^{*})$ while a proper embedding requires
${\bf 6%
}={\bf 6}^{*}$ where ${\bf 6}$=(${\bf 4}\otimes {\bf 4})_{antisym}$.,
(even
though the {\bf 6} does not enter the model). This information is
sufficient
for us to begin our investigation. We will choose $N=3$ throughout, and if we use the fact that
$SU_{L}(2)$ and $U_{Y}(1)$ are embedded in diagonal subgroups
$SU^{p}(3)$
and $SU^{q}(3)$ of the initial $SU^{n}(3)$, the ratio $\frac{\alpha
_{2}}{%
\alpha _{Y}}$ turns out to be $\frac{p}{q}.$ This implies the initial
value of $\theta_{W}$ is calculable in these models and
$\sin
^{2}\theta
_{W}$ satisfies
\be{}
\sin ^{2}\theta _{W}=\frac{3}{3+5\left( \frac{p}{q}\right) }.
\ee
On the other hand, a more standard approach is to break the initial $SU^{n}(3)$
to $SU_{C}(3)\otimes SU_{L}(3)\otimes SU_{R}(3)$ where $SU_{L}(3)$ and
$SU_{R}(3)$ are embedded in diagonal subgroups
$SU^{p}(3)$
and $SU^{q}(3)$ of the initial $SU^{n}(3)$. We then
embed all of $SU_{L}(2)$ in $SU_{L}(3)$ but $\frac{1}{3}$ of $U_{Y}(1)$ in $SU_{L}(3)$
and the other $\frac{2}{3}$ in $SU_{R}(3)$. This modifies the $\sin
^{2}\theta
_{W}$ formula to:
\be{}
\sin ^{2}\theta _{W}=\frac{3}{3+5\left( \frac{\alpha _{2}}{\alpha _{Y}}\right) }
=\frac{3}{3+5\left( \frac{3p}{p+2q}\right) }
\ee
Note, this coincides with the previous result when $p=q$. We will use the later result when calculating $\sin
^{2}\theta
_{W}$ below. A similar relation holds for Pati--Salam type models \cite{Frampton:2001xh}.

First a $Z_{2}$ orbifold has only real representations and therefore will
not
yield a chiral model. (Note, although all matter is in chiral
supermultiplet, if there is a left-handed supermultiplet to match each
right-handed supermultiplet, then the model has no overall chirality,
$i.e$., it is vectorlike.)

Next, for $\Gamma =Z_{3}$ the choice ${\bf 4}=({\bf 1},\alpha ,\alpha
,\alpha )$ with $N=3$ where $\alpha =e^{\frac{2\pi i}{3}}$ (in what
follows
we will write $\alpha =e^{\frac{2\pi i}{n}}$ for $Z_{n})$, yields the
three
family trinification \cite{Glashow:1984gc} model of \cite{Kachru:1998ys}, but without sufficient scalars
to break the gauge symmetry to the MSSM. Here the initial value of
$\sin ^{2}\theta _{W}=\frac{3}{8}$, so unification at the $TeV$ scale is also problematic.
There is another chiral model for ${\bf
4}=({\bf 1}%
,\alpha ,\alpha ,\alpha ^{2})$ but it can have at most one chiral family.

$Z_{4}$ orbifolds allow only one chiral model with ${\cal N}=1$ $SUSY.$
It
is generated by ${\bf 4}=({\bf 1},\alpha ,\alpha ,\alpha ^{2})$ but can
have
at most two chiral families.

There are two chiral models for $Z_{5}$, and they are fixed by choosing
$%
{\bf 4}=({\bf 1},\alpha ,\alpha ,\alpha ^{3})$ and ${\bf 4}=({\bf
1},\alpha,\alpha ^{2},\alpha ^{2}).$ Before looking at these in detail, let us pause to define a useful notation
for
classifying models. The initial model (before any symmetry breaking) is
completely fixed (recall we always are taking $N=3$) by the choice of
$Z_{n}$
and the embedding  ${\bf 4=}({\bf 1},\alpha ^{i},\alpha ^{j},\alpha
^{k}),$
so we define the model to by $M_{ijk}^{n}.$ We immediately observe that the
conjugate model $M_{n-i,n-j,n-k}^{n}$ contains the same information, so
we
need not study it separately.

Returning now to $Z_{5},$ the two models are $M_{113}^{5}$ and
$M_{122}^{5}.$
(Other inconsistent models are eliminated by requiring ${\bf 6}={\bf
6}^{*}$
keeping the number of models limited.) We find no pattern of spontaneous
symmetry breaking ($SSB$) for $M_{113}^{5}$ that yields the $MSSM$, but $%
M_{122}^{5}$ is more interesting. The matter content of $M_{122}^{5}$ is
shown in Table 1.

For each entry, ($\times$), in the table, we have a chiral supermultiplet
in a bifundamental representation of $SU^{5}(3)$. Specifically, for an entry
at the $i^{th}$ column and $j^{th}$ row we have a bifundamental
representation of $SU_{i}(3)\times SU_{j}(3).$ We can arbitrarily assign the
fundamental representation to the rows and the anti-fundamental
representation to the columns. If $i=j$ the bifundamental is all in
$SU_{i}(3)$ and hence is a singlet plus adjoint of $SU_{i}(3)$. Hence the
complete set of chiral supermultiplets represented by Table 1 is:
\begin{eqnarray*}
&&\lbrack (3,\bar{3},1,1,1)+(1,3,\bar{3},1,1)+(1,1,3,\bar{3},1)+(1,1,1,3,%
\bar{3})+(\bar{3},1,1,1,3)] \\
&&+2[(3,1,\bar{3},1,1)+(1,3,1,\bar{3},1)+(1,1,3,1,\bar{3})+(\bar{3}%
,1,1,3,1)+(1,\bar{3},1,1,3)] \\
&&+[(1+8,1,1,1,1)+(1,1+8,1,1,1)+(1,1,1+8,1,1)+(1,1,1,1+8,1)+(1,1,1,1,1+8)].
\end{eqnarray*}

\bigskip

\begin{tabular}{|c||c|c|c|c|c|}
\hline
$M_{122}^{5}$ & 1 & $\alpha $ & $\alpha ^{2}$ & $\alpha ^{3}$ & $\alpha
^{4}$
\\ \hline\hline
$1$ & $\times $ & $\times $ & $\times \times $ &  &  \\ \hline
$\alpha $ &  & $\times $ & $\times $ & $\times \times $ &  \\ \hline
$\alpha ^{2}$ &  &  & $\times $ & $\times $ & $\times \times $ \\ \hline

$\alpha ^{3}$ & $\times \times $ &  &  & $\times $ & $\times $ \\ \hline

$\alpha ^{4}$ & $\times $ & $\times \times $ &  &  & $\times $ \\ \hline

\end{tabular}

\bigskip

Table 1: Matter content for the model $M_{122}^{5}.$ The $\times \times $ entry at the
(1,$%
\alpha ^{2})$ position means the model contains $2(3,1,\bar{3},1,1)$ of
$%
SU^{5}(3),$ etc. The diagonal entries are $(8+1,1,1,1,1)$, etc.

\bigskip

A vacuum expectation value ($VEV$) for $(3,\bar{3},1,1,1)$ breaks the
symmetry to $SU_{D}(3)\otimes SU_{3}(3)\otimes SU_{4}(3)\otimes
SU_{5}(3)$
and a further $VEV$ for $(1,3,\bar{3},1)$ breaks the
symmetry to
$SU_{D}(3)\otimes SU_{D^{\prime }}(3)\otimes SU_{5}(3).$
Identifiny $SU_{C}(3)$ with $SU_{D}(3)$, embedding $SU_{L}(2)$ in $SU_{D^{\prime }}(3)$
and $U_{Y}(1)$ partially in $SU_{5}(3)$ and partially in $SU_{D^{\prime }}(3)$ gives an
initial value of
$\sin ^{2}\theta _{W}=\frac{2}{7}=.286$,
and implies a unification scale around
$2\times 10^{7} GeV$.

The remaining
chiral multiplets are
\be{}
3[(3,\bar{3},1)+(1,3,\bar{3})+(\bar{3},1,3)]
\ee
We have sufficient octets to continue the symmetry breaking all the
way
to $SU(3)\otimes SU(2)\otimes U(1),$ and so arrive at the MSSM with
three
families (plus additional vector-like matter that is heavy and therefore not in
the
low energy spectrum).

 Before analyzing more models in detail, it is useful to tabulate the
possible model for each value of $n$. To this end, note we always have a
proper embedding ($i.e$., ${\bf 6=6}^{*}$) for ${\bf 4=}({\bf 1},\alpha
^{i},\alpha ^{j},\alpha ^{k})$ when $i+j+k=n$. To show this we use the
fact
that the conjugate model has $i\rightarrow i^{\prime }=n-i,$
$j\rightarrow
j^{\prime }=n-j$ and $k\rightarrow k^{\prime }=n-k.$ Summing we find $%
i^{\prime }+j^{\prime }+k^{\prime }=3n-(i+j+k)=2n.$ From ${\bf
6}$=(${\bf 4}%
\otimes {\bf 4})_{antisym}$ we find ${\bf 6=}(\alpha ^{i},\alpha
^{j},\alpha
^{k},\alpha ^{j+k},\alpha ^{i+k},\alpha ^{i+j}),$ but
$i+j=n-k=k^{\prime }.$
Likewise $i+k=j^{\prime }$ and $j+k=i^{\prime }$ so ${\bf 6=}(\alpha
^{i},\alpha ^{j},\alpha ^{k},\alpha ^{i^{\prime }},\alpha ^{j^{\prime
}},\alpha ^{k^{\prime }})$ and this is ${\bf 6}^{*}$ up to an
automorphism
which is sufficient to provide vectorlike matter in this sector in the
non-SUSY models and here provide a proper embedding. Models with
$i+j+k=n$ (we will call these partition models)
are always chiral, with total chirality $\chi =3N^{2}n$ except in the
case
where $n$ is even and one of $i$, $j$, or $k$ is $n/2$ where $\chi
=2N^{2}n.$
(No more than one of $i$, $j$, and $k$ can be $n/2$ since they add to
$n$
and are all positive.) This immediately gives us a lower bound on the
number
of chiral models at fixed $n$. It is the the number of partitions of $n$
into three non-negative integers. There is another class of models with
$%
i^{\prime }=k$ and $j^{\prime }=j^{2},$ and total chirality $\chi
=N^{2}n;$
for example a $Z_{9}$ orbifold with ${\bf 4=}({\bf 1},\alpha ^{3},\alpha
^{3},\alpha ^{6}).$ And there are a few other sporadically occurring
cases
like $M_{124}^{6}$, which typically have reduced total chirality, $\chi
<3N^{2}n$.

We now tabulate all the $Z_{n}$ orbifold models up to $n=12$ along with
the
total chirality of each model, (see Table 2).

\bigskip

\bigskip
\begin{tabular}{|c||c|c|c|}
\hline
$n$ & {\bf 4} & $\chi /N^{2}$ & comment \\ \hline\hline
3 & $({\bf 1},\alpha ,\alpha ,\alpha )$ & 9 & $i+j+k=3;$ one model
$(i=j=k=1)
$ \\ \hline
3 & $({\bf 1},\alpha ,\alpha ,\alpha ^{2})^{*}$ & 3 &  \\ \hline
4 & $({\bf 1},\alpha ,\alpha ,\alpha ^{2})$ & 8 & $i+j+k=4;$ one model
\\
\hline
5 & $({\bf 1},\alpha ^{i},\alpha ^{j},\alpha ^{k})$ & 15 & $i+j+k=5;$ 2 models
\\
\hline

6 & $({\bf 1},\alpha ^{i},\alpha ^{j},\alpha ^{k})$ & 12 & $i+j+k=6;$ 3
models
\\ \hline

6 & $({\bf 1},\alpha ,\alpha ^{2},\alpha ^{4})^{*}$ & 6 &  \\ \hline
6 & $({\bf 1},\alpha ^{2},\alpha ^{2},\alpha ^{4})^{*}$ & 6 &  \\ \hline

7 & $({\bf 1},\alpha ^{i},\alpha ^{j},\alpha ^{k})$ & 21 & $i+j+k=7;$ 4
models \\ \hline
8 & $({\bf 1},\alpha ^{i},\alpha ^{j},\alpha ^{k})$ & $\leq 24$ &
$i+j+k=8;$
5 models \\ \hline
9 & $({\bf 1},\alpha ^{i},\alpha ^{j},\alpha ^{k})$ & 27 & $i+j+k=9;$ 7
models \\ \hline
9 & $({\bf 1},\alpha ,\alpha ^{4},\alpha ^{7})^{*}$ & 27 &  \\ \hline
9 & $({\bf 1},\alpha ^{3},\alpha ^{3},\alpha ^{6})^{*}$ & 9 &  \\ \hline

10 & $({\bf 1},\alpha ^{i},\alpha ^{j},\alpha ^{k})$ & 30 & $i+j+k=10;$
8
models \\ \hline
11 & $({\bf 1},\alpha ^{i},\alpha ^{j},\alpha ^{k})$ & 33 & $i+j+k=11;$
10
models \\ \hline
12 & $({\bf 1},\alpha ^{i},\alpha ^{j},\alpha ^{k})$ & $\leq 36$ &
$i+j+k=12;
$ 12 models \\ \hline
12 & $({\bf 1},\alpha ^{2},\alpha ^{4},\alpha ^{8})^{*}$ & 12 &  \\
\hline
12 & $({\bf 1},\alpha ^{4},\alpha ^{4},\alpha ^{8})^{*}$ & 12 &  \\
\hline
\end{tabular}

\bigskip

Table 2. All chiral $Z_{n}$ orbifold models with $n\leq 12.$ Three of
the $%
n=8$ models have $\chi /N^{2}=24;$ the other two have $\chi /N^{2}=16.$
Of
the 12 models with $i+j+k=12,$ three have models $\chi /N^{2}=24$ and
the
other nine have $\chi /N^{2}=36.$  Of the 60 models 53 are partition models, while the
remaining 7 models that do not satisfy $i+j+k=n$, are marked with an asterisk (*).

\bigskip

We have analyzed all the models for $Z_{6}$ orbifolds, and find only one
of
phenomenological interest. It is $M_{123}^{6}$ , where the matter
multiplets
are shown in Table 3.

\bigskip

\bigskip
\begin{tabular}{|c||c|c|c|c|c|c|}
\hline
$M_{123}^{6}$ & 1 & $\alpha $ & $\alpha ^{2}$ & $\alpha ^{3}$ & $\alpha
^{4}$
& $\alpha ^{5}$ \\ \hline\hline
1 & $\times $ & $\times $ & $\times $ & $\times $ &  &  \\ \hline
$\alpha $ &  & $\times $ & $\times $ & $\times $ & $\times $ &  \\
\hline
$\alpha ^{2}$ &  &  & $\times $ & $\times $ & $\times $ & $\times $ \\
\hline
$\alpha ^{3}$ & $\times $ &  &  & $\times $ & $\times $ & $\times $ \\
\hline
$\alpha ^{4}$ & $\times $ & $\times $ &  &  & $\times $ & $\times $ \\
\hline
$\alpha ^{5}$ & $\times $ & $\times $ & $\times $ &  &  & $\times $ \\
\hline
\end{tabular}

\bigskip Table 3: Chiral supermultiplets for the model $M_{123}^{6}.$

\bigskip

$VEV$s for $(3,\bar{3},1,1,1,1)$, $(1,1,3,\bar{3},1,1)$ and $(1,1,1,1,3,\bar{3})$ break the
symmetry to $SU_{12}(3)\otimes SU_{34}(3)\otimes SU_{56}(3)\ $where $%
SU_{12}(3)$ is the diagonal subgroup of $SU_{1}(3)\otimes SU_{2}(3),$
etc.,
and the remaining chirality resides in
$3[(3,\bar{3},1)+(1,3,\bar{3})+(\bar{3%
},1,3)]$. Again, we have octets of all six initial $SU(3)$s, so we can
break
to a three-family $MSSM$, but with $\sin ^{2}\theta _{W}=\frac{3}{8}$. There is no other pattern
of $SSB$ that gives three families.

\bigskip

For $Z_{7}$, we again find only one model that can break to a
three-family $%
MSSM$. It is $M_{133}^{7}$, with matter shown in Table 4.

\bigskip

\bigskip
\begin{tabular}{|c||c|c|c|c|c|c|c|}
\hline
$M_{133}^{7}$ & 1 & $\alpha $ & $\alpha ^{2}$ & $\alpha ^{3}$ & $\alpha
^{4}$
& $\alpha ^{5}$ & $\alpha ^{6}$ \\ \hline\hline
1 & $\times $ & $\times $ &  & $\times \times $ &  &  &  \\ \hline
$\alpha $ &  & $\times $ & $\times $ &  & $\times \times $ &  &  \\
\hline
$\alpha ^{2}$ &  &  & $\times $ & $\times $ &  & $\times \times $ &  \\
\hline
$\alpha ^{3}$ &  &  &  & $\times $ & $\times $ &  & $\times \times $ \\
\hline
$\alpha ^{4}$ & $\times \times $ &  &  &  & $\times $ & $\times $ &  \\
\hline
$\alpha ^{5}$ &  & $\times \times $ &  &  &  & $\times $ & $\times $ \\
\hline
$\alpha ^{6}$ & $\times $ &  & $\times \times $ &  &  &  & $\times $ \\
\hline
\end{tabular}

\bigskip Table 4: Chiral supermultiplets for the model $M_{133}^{7}.$

\bigskip

First $VEV$s for $(3,\bar{3},1,1,1,1,1)$ and
$(1,1,3,\bar{3},1,1,1)$ breaks
the symmetry to $SU_{12}(3)\otimes SU_{34}(3)\otimes SU_{5}(3)\otimes
SU_{6}(3)\otimes SU_{7}(3)$. Then a $VEV$ for $(1,1,1,3,\bar{3})$ breaks
this to $SU_{12}(3)\otimes SU_{34}(3)\otimes SU_{5}(3)\otimes
SU_{67}(3)$,
and leaves the following multiplets chiral
\be{}
(3,\bar{3},1,1)+(1,3,\bar{3},1)+(1,1,3,\bar{3})+(\bar{3},1,1,3)+2[(3,\bar{3}%
,1,1)+(1,3,1,\bar{3})+(\bar{3},1,1,3)]
\ee
 Finally, a $VEV$ for $(1,3,\bar{3},1)$ yields the $MSSM$ with three
chiral
families. Identifying $SU_{C}(3)$ with $SU_{12}(3)$ and embedding $SU_{L}(2)$ in $SU_{67}(3)$ and
$U_{Y}(1)$ in $SU_{345}(3)$ gives $\sin ^{2}\theta _{W}=\frac{7}{22}=.318$ and implies a
unification scale around
$ 10^{10} GeV$.

\bigskip

The $n>7$ models can be analyzed in a similar manner. The total number
of
models grows with $n$. There are also potentially interesting examples
for $%
N>4$. Although we have not made a systematic study of the models with
$n\geq %
8$, we close with a rather compact example of a three-family $MSSM$ at
$n=9$%
. The model is $M_{123}^{6}$ with matter given in Table 5.

\bigskip

\begin{tabular}{|c||c|c|c|c|c|c|c|c|c|}
\hline
$M_{333}^{9}$ & 1 & $\alpha $ & $\alpha ^{2}$ & $\alpha ^{3}$ & $\alpha
^{4}$
& $\alpha ^{5}$ & $\alpha ^{6}$ & $\alpha ^{7}$ & $\alpha ^{8}$ \\
\hline\hline
1 & $\times $ &  &  & $\times \times \times $ &  &  &  &  &  \\ \hline
$\alpha $ &  & $\times $ &  &  & $\times \times \times $ &  &  &  &  \\
\hline
$\alpha ^{2}$ &  &  & $\times $ &  &  & $\times \times \times $ &  &  &
\\
\hline
$\alpha ^{3}$ &  &  &  & $\times $ &  &  & $\times \times \times $ &  &
\\
\hline
$\alpha ^{4}$ &  &  &  &  & $\times $ &  &  & $\times \times \times $ &
\\
\hline
$\alpha ^{5}$ &  &  &  &  &  & $\times $ &  &  & $\times \times \times $
\\
\hline
$\alpha ^{6}$ & $\times \times \times $ &  &  &  &  &  & $\times $ &  &
\\
\hline
$\alpha ^{7}$ &  & $\times \times \times $ &  &  &  &  &  & $\times $  &  \\
\hline
$\alpha ^{8}$ &  &  & $\times \times \times $ &  &  &  &  &  & $\times $ \\
\hline
\end{tabular}

\bigskip

\bigskip

Table 5: Chiral supermultiplets for the model $M_{333}^{9}.$

\bigskip

$VEV$s for the octets of $SU_{k}(3)$, where $k=2,3,5,6,8,$ and $9$
breaks
the symmetry to $SU_{1}(3)\otimes SU_{4}(3)\otimes SU_{7}(3).$ (Each
chiral
supermultiplet of representation $R$ contains one chiral fermion
multiplet
in representation $R$, and two scalar (we need not distinguish scalars
from
pseudoscalars here) multiplets in representation $R$. Therefore, there
are
two scalar octets for each $SU_{k}(3)$. When one octet of $SU_{k}(3)$ is
given a $VEV$, gauge freedom can be used to diagonalize that $VEV$.
However,
there is not enough gauge freedom left to diagonalize the $VEV$ of the
second octet of the same $SU_{k}(3)$. Therefore $SU_{k}(3)$ can be
broken
completely by $SU_{k}(3)$s for the two octets). The chirality remaining
after this octet breaking is
$3[(3,\bar{3},1)+(1,3,\bar{3})+(\bar{3},1,3)]$.
Further symmetry breaking via single octets of $SU_{3}(3)$ and
$SU_{7}(3)$
leads us to the three-family $MSSM$, but with $\sin ^{2}\theta _{W}=\frac{3}{8}$.
Note that any model of the type
$M_{%
\frac{n}{3}\frac{n}{3}\frac{n}{3}}^{n}$ can be handled this way, and can
lead to a three-family $MSSM$. Hence this provides an infinite class of
three-family models.

We have found 60 chiral $Z_{n}$  orbifolds for $n\leq 12$. A systematic
search up through $n=7$ yields four models that can result in
three-family
minimal supersymmetric standard models. They are $M_{111}^{3}$,
$M_{122}^{5},
$ $M_{123}^{6},$ and $M_{133}^{7}$. We suspect there are many more
models
with sensible phenomenology at larger $n$, and we have pointed out one
example $M_{333}^{9}$, which is particularly simple in its spontaneous
symmetry breaking, and is also a member of an infinite series of models
$M_{%
\frac{n}{3}\frac{n}{3}\frac{n}{3}}^{n}$, which all can lead to
three-family
$MSSM$s.
Orbifolded $AdS/CFT$ models hold great promise for testing string theory not far above the
the $TeV$ scale, and they have inspired models \cite{Aldazabal:2000sa} with phenomenology ranging from light magnetic
monopoles \cite{Kephart:2001ix} to an anomalous muon magnetic moment\cite{Kephart:2001iu}. They have also provided a check on higher loop $\beta$
functions \cite{Pickering:2001aq}, and raised interesting cosmological questions \cite{Kakushadze:2000mc}.

\end{document}